\documentclass{article} 
\usepackage{iclr2023_conference_tinypaper,times}

\usepackage{amsmath,amsfonts,bm}

\def\eqref#1{equation~\ref{#1}}

\def\1{\bm{1}}

\DeclareMathAlphabet{\mathsfit}{\encodingdefault}{\sfdefault}{m}{sl}
\SetMathAlphabet{\mathsfit}{bold}{\encodingdefault}{\sfdefault}{bx}{n}

\usepackage{hyperref}
\usepackage{url}
\title{Connect the dots: Dataset Condensation, Differential Privacy, \& Adversarial Uncertainty}
\author{Kenneth Odoh  \\
\url{https://kenluck2001.github.io}\\
}

\iclrfinalcopy 
\begin{document}
\maketitle

\begin{abstract}

Our work focuses on understanding the underpinning mechanism of dataset condensation by drawing connections with ($\epsilon$, $\delta$)-differential privacy where the optimal noise, $\epsilon$, is chosen by adversarial uncertainty~\cite{Grining2017}. We can answer the question about the inner workings of the dataset condensation procedure. Previous work~\cite{dong2022} proved the link between dataset condensation (DC) and ($\epsilon$, $\delta$)-differential privacy. However, it is unclear from existing works on ablating DC to obtain a lower-bound estimate of $\epsilon$ that will suffice for creating high-fidelity synthetic data. We suggest that adversarial uncertainty is the most appropriate method to achieve an optimal noise level, $\epsilon$. As part of the internal dynamics of dataset condensation, we adopt a satisfactory scheme for noise estimation that guarantees high-fidelity data while providing privacy.

\end{abstract}

\section{Introduction}
We present a way of dissecting the internal dynamics of dataset condensation (DC) in connection with differential privacy (DP) and adversarial uncertainty. Dataset condensation in relation to ($\epsilon$, $\delta$)-differential privacy where we anticipate that $\epsilon$ estimated with an adversarial uncertainty scheme provides a lower-bound estimate of the noise level, $\epsilon$ resulting in a lesser impact on utility measures.

In this section, we have described the following concepts: Dataset Condensation in Subsection~\ref{data-condensation}, Differential Privacy in Subsection~\ref{diff-Privacy}, and Adversarial Uncertainty in Subsection~\ref{Adv-Uncertain}.

\subsection{Dataset Condensation}
\label{data-condensation}
Dataset condensation~\cite{dong2022} is a principled approach to creating synthetic data for machine learning ML models while retaining the original data distribution with minimal mismatch. We have explored the deep connections between DC, differential privacy, and adversarial uncertainty.

We provide a mathematical representation of dataset condensation~\cite{zheng2023} in Equation~\ref{eqn:datasetcondensation}.

\begin{equation}
\label{eqn:datasetcondensation}
\arg \min \mathbb{E}_{(x, y) \sim \tau} \ell\left(f_{\theta(\mathcal{S})}(x), y\right), \text { where } \theta(\mathcal{S})=\underset{\theta}{\operatorname{argmin}} \mathbb{E}_{(x, y) \sim \mathcal{S}} \ell\left(f_\theta(x), y\right),|\mathcal{S}| \ll|\mathcal{T}|
\end{equation}

Given data, x, with label, y, the model parameter is $\theta$, cross-entropy loss function is $\ell\left(f_\theta(x), y\right)$, synthetic data is $\mathcal{S}$, and original dataset is $\mathcal{T}$. Inferring the model parameter $\theta$ from the synthetic data, $\mathcal{S}$ to build an ML model with minimal distribution matching loss relative to the original (non-synthetic) data. Furthermore, in case there is data compression as a result of the DC algorithm, then $|\mathcal{S}| < |\mathcal{T}|$.

\subsection{Differential Privacy}
\label{diff-Privacy}

By randomizing data, this scheme obfuscates data and facilitates public release without compromising individual privacy.

\textbf{Definition 1}: (Differential Privacy) Following Definition 7~\cite{Dwork2017}, for every pair of the datasets $D$ and $D^{\prime}$, attacker's guessing advantage, $\delta$, and a randomizer, $\mathcal{M}$ satisfies $\mathrm{P}(\mathcal{M}(D) \in \mathcal{O}) \leq e^\epsilon \mathrm{P}\left(\mathcal{M}\left(D^{\prime}\right) \in \mathcal{O}\right)+ \delta$

Using the randomizer, $\mathcal{M}$, as DC method and noise, $\epsilon$. We have satisfied the ($\epsilon$, $\delta$)-differential privacy requirement DC freely provides out of the box.

\subsection{Adversarial Uncertainty}
\label{Adv-Uncertain}
The anonymization routine centered on a noiseless differential privacy scheme that considers the variance in the data as part of the noise required for randomization. We can use this idea to provide a lower-bound estimate of noise. There is no point in adding more noise than necessary since the data already contain inherent noise, and they will have a lesser impact on utility if there is less added noise.

Adversary uncertainty supports both the default threat model in Theorem 1 and a relaxed threat model in Theorem 2 (see appendix). The default threat model provides privacy protection where the adversary does not know the original data before randomization. This assumption can be problematic, as the data can be re-identified by correlation with ancillary data after public release. On the contrary, the relaxed threat model does not compromise privacy, even if the adversary knows the proportion of the original data.

\section{Discussion}
Previous works have shown the relationship between DC and ($\epsilon$, $\delta$)-DP scheme~\cite{dong2022} that provides needed privacy protection without extra processing by default. Synthetic data created using DC have high quality while preserving indistinguishability between generated data samples, thereby maintaining the privacy of individual records. When $\epsilon$ is estimated using adversarial uncertainty, a lower-bound noise estimation can be used without negatively impacting its utility. 

Synthetic data are beneficial for training ML models. We can empirically infer the value of $\epsilon$~\cite{dong2022} from the output of a DC routine. Hence, we have shown the relationship between DP and DC. ~\cite{kairouz2015} has proved the relationship between false positive (FP), false negative (FN), an attacker advantage, $\delta$ in $(\epsilon, \delta)$-differentially private mechanism, $\mathcal{M}$ as shown in Equations~\ref{eqn:fp},~\ref{eqn:fn} (classification model).

\noindent\begin{minipage}{.5\linewidth}

\begin{equation}
\label{eqn:fp}
FP+e^{\epsilon} FN \leq 1-\delta
\end{equation}
\end{minipage}%
\begin{minipage}{.5\linewidth}
\begin{equation}
\label{eqn:fn}
FN+e^{\epsilon} FP \leq 1-\delta
\end{equation}
\end{minipage}

Reduce Equations~\ref{eqn:fp},~\ref{eqn:fn} to Equations~\ref{eqn:fpres},~\ref{eqn:fnres} respectively.

\noindent\begin{minipage}{.5\linewidth}

\begin{equation}
\label{eqn:fpres}
\epsilon \leq \ln{ \left( \frac{1-\delta-FP}{FN}\right) } 
\end{equation}
\end{minipage}%
\begin{minipage}{.5\linewidth}
\begin{equation}
\label{eqn:fnres}
\epsilon \leq \ln{ \left( \frac{1-\delta-FN}{FP}\right) } 
\end{equation}
\end{minipage}

In addition, the worst-case noise, $\epsilon$, can be estimated by Equation~\ref{eqn:noise} by taking the maximum noise in Equations~\ref{eqn:fpres},~\ref{eqn:fnres}.
\begin{equation}
\label{eqn:noise}
\epsilon = \max \left(\ln{ \left( \frac{1-\delta-FP}{FN}\right) }, \ln{ \left( \frac{1-\delta-FN}{FP}\right)} \right)
\end{equation}

We can extend this work by considering an example where a mixture of synthetic data from the dataset condensation and the original data. Adversarial uncertainty permits a relaxed threat model to provide privacy protection, even if the attacker has seen only a fraction of the original data. This scenario is popular due to the widespread practice of using a pre-trained model based on a foundation model. Most large-language models that utilize training data scraped from the internet may suffer a high risk of data contamination.

\section{Conclusion}
We have shown a strong link between dataset condensation, differential privacy, and adversarial uncertainty. To make sense of our contribution, we can recreate the scheme where the randomizer function is the dataset condensation routine satisfying the ($\epsilon$, $\delta$)-differential privacy, given the need for a lower-bound estimate of $\epsilon$, hence, the need for adversarial uncertainty to calculate the noise, $\epsilon$, thereby minimizing the impact on utility.

\subsubsection*{URM Statement}

The authors acknowledge that at least one key author of this work meets the URM criteria of ICLR 2024 Tiny Papers Track.

\bibliography{iclr2023_conference_tinypaper}

\begin{thebibliography}{5}
\providecommand{\natexlab}[1]{#1}
\providecommand{\url}[1]{\texttt{#1}}
\expandafter\ifx\csname urlstyle\endcsname\relax
  \providecommand{\doi}[1]{doi: #1}\else
  \providecommand{\doi}{doi: \begingroup \urlstyle{rm}\Url}\fi

\bibitem[Dong et~al.(2022)Dong, Zhao, and Lyu]{dong2022}
Tian Dong, Bo~Zhao, and Lingjuan Lyu.
\newblock {Privacy for Free: How does Dataset Condensation Help Privacy?}
\newblock In \emph{{Proceedings of the 39th International Conference on Machine
  Learning}}, 2022.

\bibitem[Dwork et~al.(2017)Dwork, Smith, Steinke, and Ullman]{Dwork2017}
Cynthia Dwork, Adam~D. Smith, Thomas Steinke, and Jonathan Ullman.
\newblock {Exposed! A Survey of Attacks on Private Data}.
\newblock \emph{Annual Review of Statistics and Its Application}, 4:\penalty0
  61--84, 2017.

\bibitem[Grining \& Klonowski(2017)Grining and Klonowski]{Grining2017}
Krzysztof Grining and Marek Klonowski.
\newblock {Towards Extending Noiseless Privacy: Dependent Data and More
  Practical Approach}.
\newblock In \emph{{Proceedings of the 12th Asia Conference on Computer and
  Communications Security}}, pp.\  546--560, 2017.

\bibitem[Kairouz et~al.(2015)Kairouz, Oh, and Viswanath]{kairouz2015}
Peter Kairouz, Sewoong Oh, and Pramod Viswanath.
\newblock {The Composition Theorem for Differential Privacy}.
\newblock In \emph{{Proceedings of the 32nd International Conference on Machine
  Learning}}, pp.\  1376--1385, 2015.

\bibitem[Zheng \& Li(2023)Zheng and Li]{zheng2023}
Tianhang Zheng and Baochun Li.
\newblock {Differentially Private Dataset Condensation}, 2023.
\newblock URL \url{https://openreview.net/forum?id=H8XpqEkbua_}.

\end{thebibliography}
\bibliographystyle{iclr2023_conference_tinypaper}


\section{Appendix}
\textbf{Theorem 1} (Following Theorem 2 of~\cite{Grining2017}): Given that data vector, $X=\left(X_1, \ldots, X_n\right)$ where each $X_i$ is an independent random variable. Fix $\mu_i=E X_i$ and $\sigma^2=$ $\frac{\sum_{i=1}^n \operatorname{Var}\left(X_i\right)}{n}$ and $E\left|X_i\right|^3<\infty$ for every $i \in\{1, \ldots, n\}$.

We denote a given mechanism $M(X)=\sum_{i=1}^n\left(X_i\right)$ and data sensitivity as $\Delta$. $M(X)$ is $(\epsilon, \delta)-DP$ with the following parameters:
$$
\epsilon=\sqrt{\frac{\Delta^2 \ln (n)}{n \sigma^2}}
$$
And
$$
\delta=\frac{1.12 \sum_{i=1}^n E\left|X_i-\mu_i\right|^3}{\left(n \sigma^2\right)^{\frac{3}{2}}}\left(1+e^\epsilon\right)+\frac{4}{5 \sqrt{n}}.
$$
\textbf{Theorem 2} (Following Theorem 5 of~\cite{Grining2017}): Given that data vector, $X=\left(X_1, \ldots, X_n\right)$ where each $X_i$ is an independent random variable. Let us assume that the attacker knows the original values of a proportion, $\gamma$ of users. Denote the set of indexes of compromised users by $\Gamma$, where $|\Gamma|=\gamma n$. Using the mechanism $M(X)=\sum_{i=1}^n\left(X_i\right)$. Let $E X_i=\mu_i$ and $E X_i^4<\infty$. Let's fix $\sigma_{\Gamma}^2=\operatorname{Var}(X \backslash \Gamma)$, and data sensitivity is $\Delta$ then $M(X)$ is $(\epsilon, \delta)$-DP with following parameters
$$
\epsilon=\sqrt{\frac{\Delta^2 \ln ((1-\gamma) n)}{\sigma_{\Gamma}^2}}
$$
And
$$
\delta=c(\epsilon) \sqrt{\frac{D^2}{\sigma_{\Gamma}^3} M_X^3+\frac{D^{\frac{3}{2}} \sqrt{26}}{\sigma^2 \sqrt{\pi}} \sqrt{M_X^4}}+\frac{4}{5 \sqrt{(1-\gamma) n}},
$$
Where
$$
\begin{aligned}
M_X^3 & =\sum_{i \in[n] \backslash \Gamma} E\left|X_i-\mu_i\right|^3 \\
M_X^4 & =\sum_{i \in[n] \backslash \Gamma} E\left(X_i-\mu_i\right)^4
\end{aligned}
$$
And
$$
c(\epsilon)=2\left(1+e^\epsilon\right)\left(\frac{2}{\pi}\right)^{\frac{1}{4}} .
$$

\end{document}